\documentclass[11pt]{article}
\usepackage{graphicx} 
\usepackage{blois,epsfig}
\usepackage[latin1]{inputenc}
\usepackage[OT1]{fontenc}

\def\be{\begin{equation}}
\def\ee{\end{equation}}
\def\bea{\begin{eqnarray}}
\def\eea{\end{eqnarray}}

\begin{document}
\vspace*{4cm}
\title{ARE AGN THE BEST FACTORIES \\ FOR HIGH ENERGY PARTICLES AND PHOTONS? 
}

\author{ SUZY COLLIN}

\address{LUTH, Observatoire de Paris, CNRS, Universit\'e Paris Diderot; 5 Place Jules Janssen, 92190 Meudon, France}

\maketitle
\abstracts{
The main properties of AGN are reviewed, focussing on the accretion process and on the question of whether AGN are the best factories of ultra high energy particles and photons. I recall the large differences between the accretion/ejection flows in strong and weak accretors, and I conclude that, since low luminosity AGN and even ``dormant" massive black holes in nuclei of galaxies are powering strong confined magnetized jets able to accelerate high energy particles, and are present in a large proportion of galaxies, they might be better potential sources of high energy particles and photons than luminous AGN and powerful radio galaxies. }

\section{Introduction}

The recent discovery of a correlation between the directions of ultra high energy cosmic rays
(UHECRs) with Active Galactic Nuclei (AGN)
closer than 75 Mpc \cite{auger} triggered a controversy about the origin of these events. Gorbunov et al. \cite {Gorbunov} argue that
the AGN hypothesis is not supported by the data, because there is a deficit of events from the direction of the Virgo
supercluster.
Kotera \& Lemoine \cite {kotera}   show that the observed positions must correspond to inhomogeneities in the last scattering surface of UHECRs and not to the source population. Here I shall adopt the assumption that the inhomogeneities in the spatial distribution of UHECR are due to the source population, either luminous AGN, or objects having a similar spatial distribution. Quite few possibilities have been envisioned so far: AGN, Gamma Ray bursts, and shocks in galaxy clusters. I will reiterate here another proposal, which was already made by Boldt \& Ghosh in 1999 \cite{Boldt}.

AGN are bright nuclei of galaxy whose luminosity is too important to be provided by  stellar light.  Besides their high luminosity, they present common features, such as a broad band spectrum extending on many decades of energy, in particular in the X-ray range, and a strong variability.  They have several other properties which they do not all share, like intense and broad spectral lines, a high degree of polarization, or a radio and a gamma-ray emission.
 It is now universally admitted that they are powered by accretion onto a supermassive black hole (cf. Rees ~\cite{Rees}). In luminous AGN, one thinks that accretion takes place via an ``accretion disc". The properties of the  disc - its structure, dynamics, emission - depend in particular on the accretion rate and on the mass of the black hole. 
 
 Initially, only quasars and Seyfert were considered as being AGN, but it was rapidly recognized that many other galactic nuclei possess massive black holes, sometimes accreting at a quite low rate. Presentlly, one thinks that almost each nucleus of galaxy contains a black hole, and that about 40$\%$ of them are active at some level (Ho \cite{Ho99}, \cite{Ho02}). 

The ``Unified Scheme" recognizes the importance of the orientation. It was discovered for Seyfert nuclei, was later extended to other active objects, and helped to understand their properties. But an important problem remains: some objects are radiating strongly in the radio range (they are called ``radio-loud" AGN), others - the majority - are not (``radio-quiet" AGN). These classes differ by some important properties, leading to the idea of a real dichotomy between them, which is not well-understood.

Since the audience here consists mainly of astro-particle physicists, I recall the general properties of AGN in the next section. The accretion/ejection mechanisms of high luminosity AGN are discussed in Section 3, and those of low luminosity ones in Section 4, as well as tentative conclusions.

\section{General properties of AGN}

AGN span a large range of luminosity, extending from quasars (10$^{46}$-10$^{48}$ ergs s$^{-1}$), to Seyfert nuclei (10$^{43}$-10$^{46}$ ergs s$^{-1}$). ``Low Luminosity AGN" (LLAGN, 10$^{40}$-10$^{43}$ ergs s$^{-1}$) were recently also included in the AGN class. The luminosity here is actually the ``bolometric luminosity" which takes into account the whole energy spectrum from radio to gamma-rays. The separation between quasars and Seyfert galaxies is artificial, since the host galaxies of many quasars are now detected up to redshifts of 2-3. 

Supermassive Black Holes (SMBHs) at the center of galactic nuclei span also a large range of masses, from 10$^{10}$M$_{\odot}$ in the most luminous quasars, to 10$^{5}$M$_{\odot}$ in Low Luminosity AGN. Very few - and dubious - masses between 100M$_{\odot}$ and 10$^{5}$M$_{\odot}$ have been found so far,  raising the question whether they exist or not.  SMBHs are probably present in all nuclei of galaxy, and their masses are related to the luminosity of the bulge of the host galaxy (Magorrian et al. \cite{Magorrian}). The ``bulge" is a spheroidal region present in all galaxies except irregular ones, whose size and luminosity depends on the Hubble type: they decrease from early elliptical galaxies (which consist almost entirely in a massive bulge) to late spiral galaxies.  Another better relationship (Gebhardt et al. \cite{Gebhardt}, Ferrarese \& Meritt \cite{Ferrarese}) was found between the masses of the SMBHs and those of the bulges \footnote{Actually the dispersion velocity in the bulge, which is directly related to its mass.}.  It was also shown that AGN and quasars, when their host galaxy can be observed, follow about the same relationship, proving that there is nothing fundamentally different in the black hole mass of active versus quiescent galactic nuclei. The black hole masses are roughly equal to 2/1000 of the bulge masses, and the reason of this relationship which links the formation and growth of black holes to those of the whole galaxy is still not well understood.

In the local Universe, quasars are absent \footnote{Because they were more numerous and luminous in the past.} and luminous Seyfert galaxies constitute only 1$\%$ of all galaxies. Since a majority of galaxies possesses a nucleus containing a SMBH, we deduce that there are dormant black holes. But actually a low level of activity is maintained in about 40$\%$ of them, divided in weak Seyfert nuclei and in LINERs (``Low Ionization Narrow Emission Line Regions"), an heterogeneous class of galactic nuclei whose optical-UV continuum is undoubtedly non-stellar. All these objects are referred to as Low Luminosity AGN (LLAGN). Their observational properties are different from normal AGN, as we shall see later on.

Luminous AGN take different aspects, which led to distinguish several classes between them. Not only are they radio-loud or radio-quiet, but  also ``Optically Violently Variable", ``Highly Polarized", etc... ``BL Lac" are strongly variable radio-loud objects whose spectrum is devoid of emission lines. ``Blazars" are very similar, but display emission lines. Seyfert nuclei are themselves divided into type 1 (with broad {\it and} narrow emission lines), and type 2 (with {\it only} narrow lines). 
The ``Unified Scheme" of Antonucci \& Miller \cite{Antonucci} allowed to understand these differences.  
They found broad emission
lines in the polarized light of a famous Seyfert 2 galaxy, NGC 1068. They deduced  that Seyfert 2
have also broad spectral lines, but these lines are absorbed in some directions by a dusty torus surrounding the emission region - in a sense the prolongation of the accretion disc - and are scattered into our line of sight, probably by a hot - and therefore polarizing - medium. If we observe a Seyfert 1 in natural light  through the torus, it will have no broad lines and will appear as a Seyfert 2, while in polarized light we will see the reflected broad lines.
This discovery brought into focus that orientation is an important factor.

Radio-galaxies were first not considered as AGN, and indeed they do not satisfy the definition of being ``bright" galactic nuclei. They were divided in two classes by Fanaroff and Riley \cite{Fanaroff}. FRI are faint and characterized by bright jets, not highly relativistic, extending far from the center (up to hundred thousands of parsecs) and decelerated during their propagation (good examples are M87 and Centaurus A),  while FRII are powerful and have faint relativistic jets which remain  relativistic all the way up to intense ``hot spots" located at the end of big radio lobes (like Cygnus A). The division between FRI and FRII  might depend on the luminosity and on the mass of the host-galaxies, and it is also proposed that FRI objects are fuelled by accretion
from hot gas (present in elliptical galaxies and in the intergalactic medium at the center of clusters), and FRII from cold gas. Since FRII and radio-loud quasars have very similar radio morphology,  Barthel~\cite{Barthel} proposed that FRIIs are actually radio-loud quasars seen through an obscuring torus which absorbs the central optical-ultraviolet continuum and the broad lines. This idea received several confirmations, and there is no doubt now that powerful FRIIs are hiding quasars in their core. Moreover, if a radio-loud quasar is seen along the jet axis, it will appear as a blazar, owing to Doppler
boosting. 
 On the other hand, FRIs share common properties with BL Lac
objects, and their statistical properties with those of their host galaxies agree with the hypothesis of FRIs being the parent - i.e. the not relativistically boosted - population of BL Lacs (Urry \& Padovani \cite{Urry}). From the properties of the boosted and non-boosted populations, one deduce that the bulk Lorentz factors of the jets are of the order of 10 at most, but the relativistic electrons giving rise to the synchrotron radiation have random Lorentz factors up to 10$^6$.

\begin{figure}
\begin{center}
%\epsfxsize=10cm 
%\epsfbox{SU.eps}
\includegraphics[width=10cm]{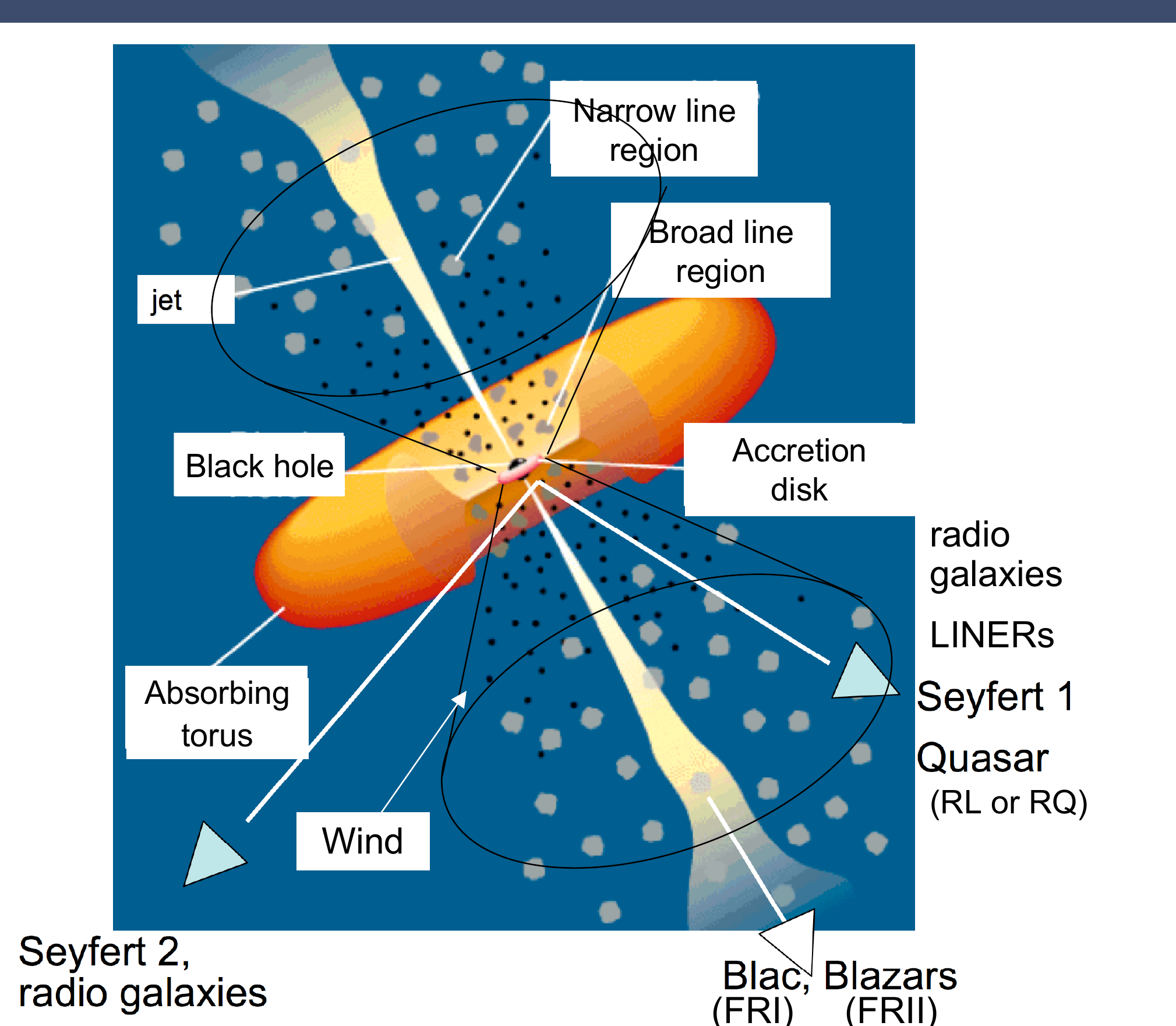}
\caption{Figure summarizing the Unified Scheme, adapted from Urry \& Padovani 1995.}
\label{su}
\end{center}
\end{figure}

Fig. \ref{su} illustrates the Unified Scheme.

\begin{figure}
\begin{center}
%\epsfxsize=15cm 
%\epsfbox{sikora.eps}
\includegraphics[width=15cm]{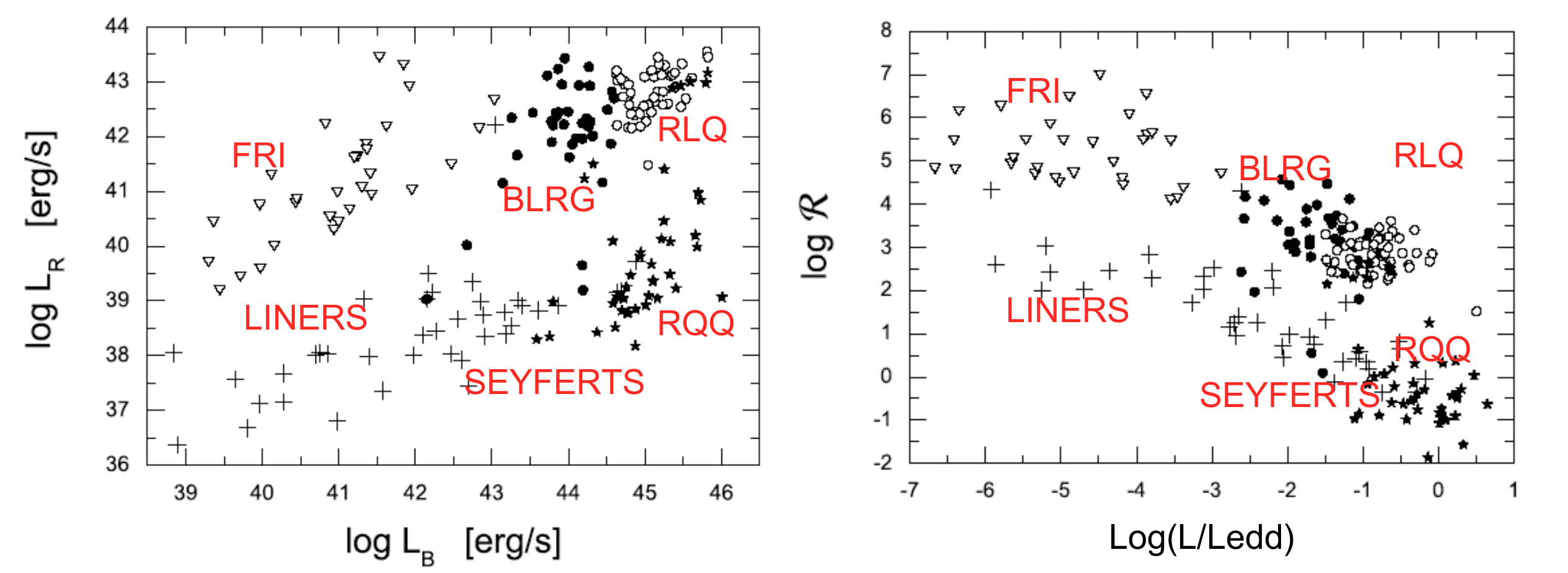}
\caption{Left: 5 GHz luminosity versus the B-band nuclear luminosity, after Sikora et al. 2007. Right:  radio loudness versus the Eddington ratio. The horizontal line separate radio-loud from radio-quiet objects. The names of the different classes of AGN are given (see the text for explanations). }
\label{sikora}
\end{center}
\end{figure}

AGN are divided traditionally into radio-quiet and radio-loud objects, whose radio luminosity differs typically by three orders of magnitude for the same optical luminosity. The left panel of Fig. \ref{sikora} from Sikora et al. \cite{Sikora} shows the radio power at 5 GHz versus the luminosity in the blue, for a sample of AGN spanning a large range of luminosities and including LLAGN (blazars are suppressed from the sample). One sees clearly two parallel sequences where the optical and radio luminosities are roughly proportional. The different classes of objects already mentioned are indicated (RLQ and RQQ: respectively radio-loud and radio-quiet quasars), with the addition of a relatively rare class: ``Broad Line Radio-Galaxies" or BLRG, which have similar characteristics as Seyfert galaxies but are elliptical galaxies, while ordinary Seyferts are spiral. 
 On the right panel of Fig. \ref{sikora}, the radio-loudness, i.e.  the radio to optical luminosity ratio, $R$, is plotted as a function of the Eddington ratio. The two parallel sequences display an anti-correlation with the Eddington ratio. According to this figure, Seyfert nuclei are all radio-quiet, as well as a large fraction of LINERs. 
 
RQ and RL AGN differ by some important properties. 

\smallskip
\noindent 1 - RL emit gamma-rays, which are not observed in RQ.

\smallskip
\noindent 2 - RL  are associated with jets and RQ with winds. There are strong differences between both outflows: jets are made of relativistic particles, are collimated and are radiating non-thermally, while winds are not collimated and are made of a warm plasma ($\sim 10^5$K), with velocities of only few hundreds up to at most 20000 km s$^{-1}$. While jets are launched in the close vicinity of the BH at  $R\le 10^{2-3}R_{\rm G}$, winds are launched at larger distances.

\smallskip
\noindent 3 - RL are located exclusively inside elliptical galaxies, and RQ in spirals \footnote{This is true only in the local Universe, as galaxies loose their identity in the past.}. 

\smallskip
We will discuss the significance of these properties in the next section. 

Finally, it is worthwhile recalling some fiducial numbers. An important one is the gravitational radius $R_{\rm G}$, equal to $GM/c^2$, where $M$ is the black hole mass. It is equal to 1.5 10$^{13}M_8$ cm, or 0.5 $10^{-5}M_8$ pc, where $M_8$ is expressed in 10$^8$ M$_{\odot}$, a typical BH mass in AGN. The Eddington luminosity, $L_{\rm Edd}$, gives the maximum radiation
power due to accretion. It
corresponds to the equality between the 
gravitational force 
exerted on protons and the radiative force acting on electrons.
In a fully ionized gas,
$L_{\rm Edd}=4\pi\ c\ GM\mu_e / \sigma_{\rm T}=1.5\times 10^{46} M_8\ {\rm erg\ s}^{-1} $,
where  $\sigma_{\rm T}$ is the Thomson cross section and $\mu_e$ the unit mass per electron. Finally the bolometric luminosity $L_{\rm bol}$ is equal to $\epsilon \dot{M} c^2$, where $\epsilon$ is the maximum efficiency of mass-energy conversion, and $\dot{M}$ the mass accretion rate. 
$\epsilon$ depends on the angular momentum of the black hole. If it does not rotate (Schwarzschild BH), the Innermost Stable Circular Orbit (ISCO) is equal to 6 $R_{\rm G}$. Below this radius, the accreting gas plunges inside the BH without radiating anymore, and $\epsilon = 0.057$ \footnote{The efficiency is probably larger, as the stress  tensors are not completely cancelled inside this radius and the gas can still radiate some energy.}. For a maximally rotating BH (a Kerr BH),  the surrounding space itself is dragged into  the rotation,  the ISCO is about equal to $R_{\rm G}$, so the radiation efficiency is larger, and $\epsilon$ reaches 30$\%$. Taking $\epsilon=0.1$ as an average value, one gets the accretion rate: $\dot{M}= 0.6 (\epsilon/0.1) L_{46}$ M$_{\odot}$ yr$^{-1}$, where $L_{46}$ is the luminosity expressed in 10$^{46}$ ergs s$^{-1}$. Finally, an important parameter is the ``Eddington ratio" $L/L_{\rm Edd}$. Quasars and Seyfert galaxies are characterized by Eddington ratios between 0.01 and unity, while in LLAGN, they are lower than 0.01.
The corresponding ``Eddington accretion rate" is
$\dot{M}_{\rm Edd}={L_{\rm Edd}\over c^2\epsilon} = 2.8\ M_8\left({\epsilon/
0.1}\right)^{-1} \  {\rm M_{\odot}\ yr^{-1}} $,
and we define $\dot{m}=\dot{M}/\dot{M}_{\rm Edd}$.

\section{Strong accretors, quasars and Seyfert nuclei,  $\dot{m} \ge 0.01$}

Contrary to stellar black holes, supermassive black holes must fish for their accretion at large distances where enough matter is available. For very luminous objects (quasars), this is achieved through mergers of massive galaxies, one at least being gas rich. A simple interaction between two galaxies can induce tidal motions which transports the matter towards the nucleus at a rate sufficient for a Seyfert nucleus. For LLAGN, non-axisymmetric potentials, bars, and dynamical friction of molecular clouds, can also pull some gas towards the nucleus. 
When the gas arrives in the innermost regions located at less than 10$^3R_{\rm G}$, i.e. $\sim 10^{-2}M_8$ pc, the transport can be accomplished through a viscous accretion disc. The disc is gravitationally unstable at distances larger than 10$^3R_{\rm G}$, and it is probably destroyed. We basically ignore what happens in a ``no man's land" located between 10$^3$ and 10$^6R_{\rm G}$.

 Let us consider the accretion disc. It is widely accepted that the mechanism for transporting the
angular momentum from the black hole
can be identified with a ``turbulent viscosity", according 
to the  ``$\alpha$-prescription" proposed by Shakura \& Sunyaev \cite{Shakura}, 
which assumes that the sizes of 
the turbulent eddies are 
smaller than the 
thickness 
of the disc and that the turbulence is subsonic.
This model gives satisfying results when applied to discs in 
cataclysmic variables.
A magnetohydrodynamic instability pointed out by Chandrasekhar \cite{Chandra}
is able  
to provide the viscosity (cf. Balbus \& Hawley \cite{Balbus}). 
 Other prescriptions (e.g. based on 
purely hydrodynamic
instabilities, Dubrulle et al. \cite{Dubrulle})  are also under 
debate.
Finally the viscosity 
can be replaced by a non-local mechanism, for instance an organized magnetic 
field anchored in the disc.

\begin{figure}
\begin{center}
%\epsfxsize=10cm 
%\epsfbox{Sanders.eps}
\includegraphics[width=10cm]{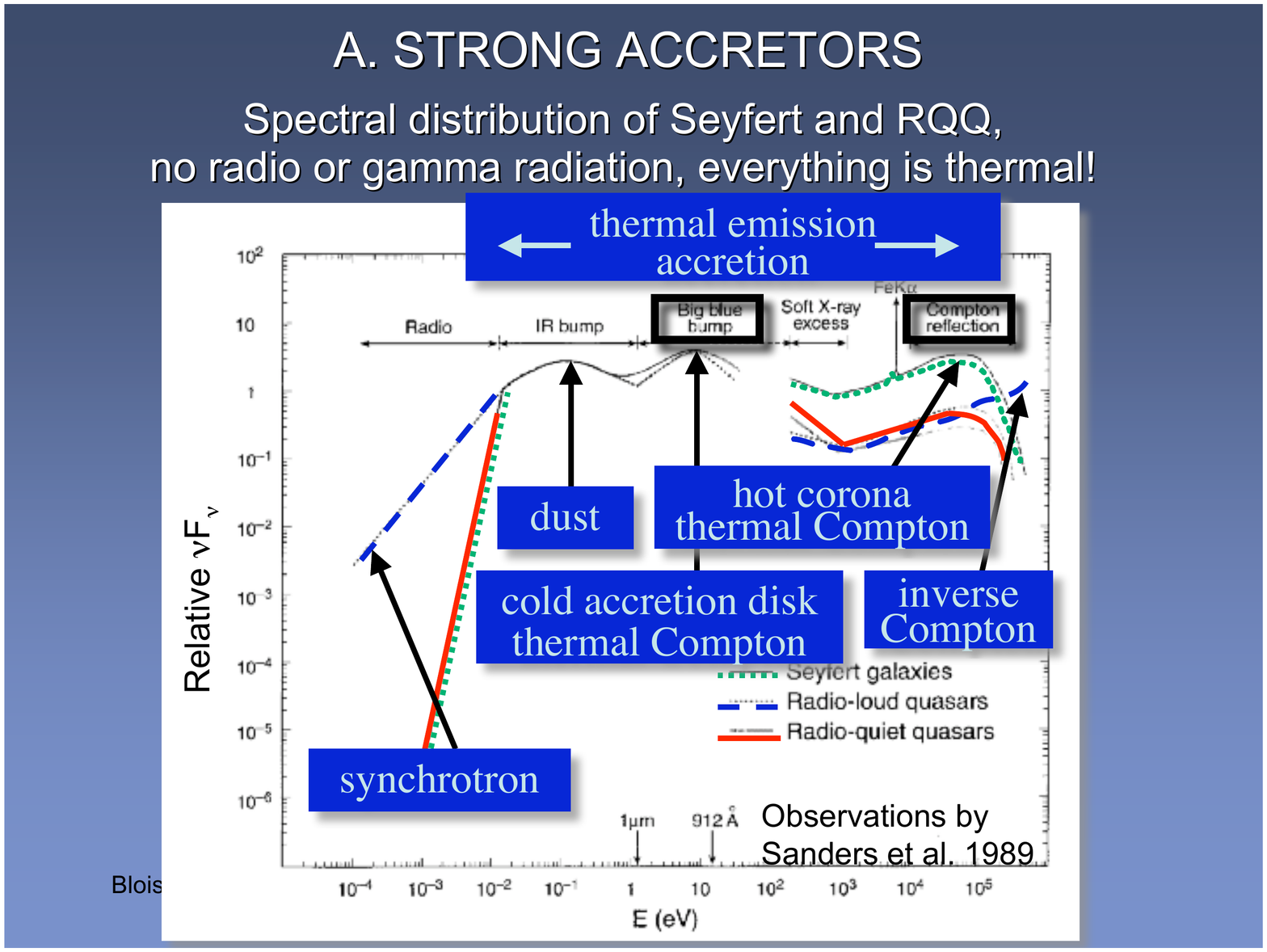}
\caption{The typical spectral energy distribution for radio quiet Seyfert nuclei and a sample of radio-loud and radio-quiet quasars, from Sanders et al. 1989. The origin of the emission is indicated, and one sees that the whole emission of radio-quiet objects is thermal.}
\label{Sanders}
\end{center}
\end{figure}

\medskip
Fig. \ref{Sanders} displays the average spectral energy distribution of Seyfert nuclei (radio-quiet) and of a sample of radio-quiet and radio-loud quasars  from Sanders et al. \cite{Sanders}. It appears immediately that only radio-loud AGN produce gamma-rays. 

\subsection{Radio-quiet AGN}

In radio-quiet AGN, the continuum is limited to the infrared at long wavelengths, and to hard X-rays at small wavelengths. It can be divided into 
three ``bumps", an ``IR-bump", a ``UV-bump", and an ``X-bump". 
 The IR-bump is 
attributed to dust (presumably in the absorbing torus) heated by the central UV-X continuum. The UV-bump, called also the ``Big Blue Bump", is due to the accretion disc. 

It is out of the scope of this talk to develop the formalism of $\alpha$-discs (see for instance Frank et al. \cite{Frank}). We mention only some important results concerning stationary discs. 

\smallskip
\noindent $0.01 \le \dot{m} \le 1$

The discs are {\it geometrically thin}, with $H/R \le 0.01$, where $R$ is the radius and $H$ the corresponding scale height. They are optically thick and emit locally like a blackbody or a modified blackbody, the bulk of the emission being in the UV band. Radial pressure gradients are negligible and the velocity is purely rotational and Keplerian. This case applies to modest quasars and to Seyfert galaxies.

\smallskip
\noindent$\dot{m} \ge 1$

What happens if the black hole is embedded in a dense medium providing more than the Eddington rate for accretion? Since the Eddington ratio cannot be much larger than unity (and is indeed observed so), $\epsilon$ should then decrease. Owing to the large accretion rate, the flow is dense and optically thick, the
inner regions are supported by a strong radiation pressure, 
consequently {\it the disk is inflated and becomes geometrically thick} \footnote{For $0.3 \le \dot{m} \le 1$
the discs are still geometrically thin, with $H/R \sim 0.1$, and are called ``slim discs" (Abramowicz et al. \cite{Abram}). Radial pressure gradients are no more negligible, and the velocity begins to depart from Keplerian. This case applies to luminous quasars and to a fraction of Seyfert 1 galaxies (called ``Narrow Line Seyfert 1"). }. Radial and rotational velocities become comparable, matter is driven very rapidly to the BH, and photons (which undergo many electron scatterings) are swallowed by the BH before they  can escape. Heat advected to the
BH exceeds thus
the heat radiated away, and
 $\epsilon$ decreases with increasing $\dot{m}$. These disks radiate 
mainly in the soft X-ray range and they could 
account for extreme NLS1s, which  display large soft X-ray bump, and for luminous quasars. However the issue is still controversial because MHD numerical simulations show that when the gas supply is super-Eddington at large distances, it is blown out before reaching the black hole (Mineshige \& Ohsuga \cite{Mineshige}).

\begin{figure}
\begin{center}
%\epsfxsize=15cm 
%\epsfbox{SpeX.eps}
\includegraphics[width=15cm]{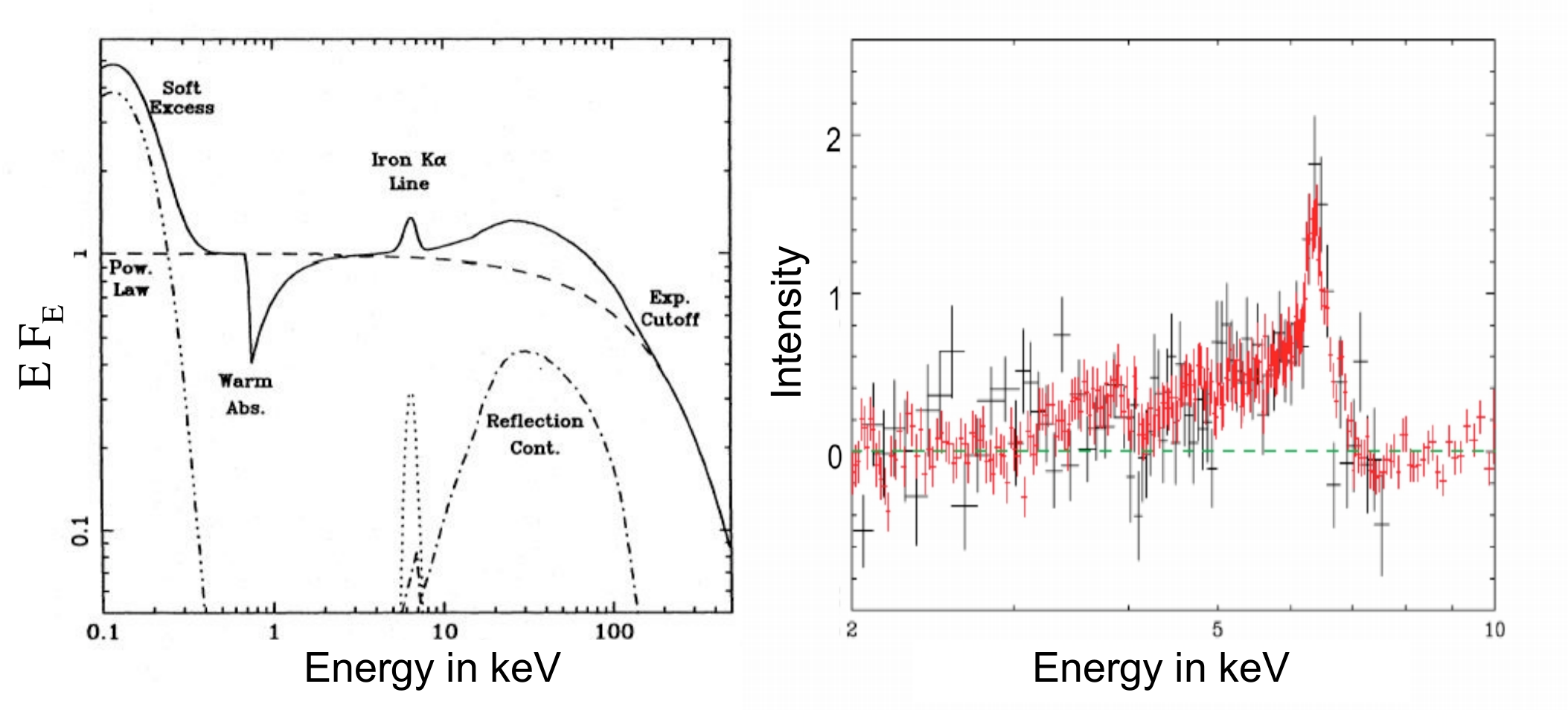}
\caption{Left:  the different components of the X-ray spectrum. Right: Profile of the FeK line in MCG -6-30-15, observed with the X-ray telescopes Chandra and XMM-Newton, showing a very extended red wing and an abrupt blue decline, from Miniutti et al. 2007.}
\label{SpeX}
\end{center}
\end{figure}

\smallskip
Standard accretion discs account for the Big Blue Bump, but do not predict  the  X-ray bump. To understand its origin, one must study in detail the X-ray spectrum. Fig. \ref{SpeX} shows  the whole X-ray spectrum of a Seyfert nucleus (left panel), and the FeK line at 6.4 keV (right panel). In  the left panel,  we see that the 
 X-ray spectrum can be decomposed in several components: a hard power-law with a turnover at a few hundreds keV, a soft X-ray excess, a ``reflection" component (continuum plus FeK line) made of backscattered X-rays due to the irradiation of a surrounding ``cold" medium, most probably the accretion disc itself, and finally an absorption due to an ionized plasma called the ``Warm Absorber". The disc itself should thus consist of two parts: 

\noindent - the ``standard" disc emitting the Big Blue Bump,

\noindent - a very hot optically thin corona surrounding the inner regions of the disc and emitting X-rays mainly by  thermal Comptonization of the ultraviolet disc photons. The corona is possibly made of a few active regions sustained by magnetic  loops anchored in the disc, whose reconnection stimulates flares, like in the solar corona. 

A key question is whether the standard disc stretches all the way to the innermost stable circular orbit, or if it is  disrupted in the inner regions and replaced by an optically thin hot flow. The shape of the iron line contains in principle the answer to this question, since the line forms only in the disc part of the flow. 
 The line displays often a peculiar profile with a broad red wing (right of Fig. \ref{SpeX}), which means that photons have undergone a strong gravitational redshift. In the case of the Seyfert nucleus MCG-6-30-15,  the disc reaches 2$R_{\rm G}$, proving that the object contains a rapidly spinning BH (Miniutti et al. \cite{Min}). Moreover, the variability properties of the X-ray flux imply a strong gravitational light bending effect, consistent with a location of the X-ray source very close to the BH. Even in this case, {\it the emission is thermal and does not extend in the gamma-ray range}.
 
 The Warm Absorber is identified with an outflowing wind surrounding the black hole and the X-ray source. It is detected by X-ray absorption features in about 50$\%$ of Seyfert nuclei, and by broad UV absorption lines in about 15$\%$ of radio-quiet quasars. This small fraction is probably simply an effect of covering factor, the wind being a common feature in radio-quiet objects. The outflowing mass rate is not well determined, but it is certainly a non-negligible fraction of the accretion rate, and it increases with the Eddington ratio. The wind is most naturally interpreted as outflowing from the disc, but there is no consensus on the  acceleration mechanism: radiative acceleration by the UV-X source, centrifugal driving from a disc threated by an open magnetic field, thermal or hydrodynamical driving from the hot corona. 
 
 \subsection{Radio-loud AGN}
 
 Radio-loud quasars and BLRGs present jets similar to those of FRII galaxies, which energize big radio lobes extending up to millions of parsecs. Reacceleration of particles take place at the extremities of the lobes in the ``hot spots" which can probably be identified as shock waves due to the jet hitting accumulated intergalactic matter. Jets at large scale transport an energy comparable to the radiation power of the nucleus. 

The radio emission is due to the synchrotron process from relativistic electrons. It is not confined to the radio range, and can extend up to the UV and even to the X-ray range. In this case,  the energy of electrons is of the order of TeVs, and if protons are associated to these electrons, their energy can reach 10$^{15}$ eV, which is still far below the highest energy of cosmic rays. Two models are presently proposed for the gamma-ray emission: 1. the leptonic models where they are produced by Inverse Compton scatterings between the relativistic electrons and either the synchrotron radio photons (``Synchrotron-Self-Compton", or SSC), or external soft photons; 2. hadronic models, which account for very high energy gamma-rays by synchrotron emission from hadrons, or interaction of hadrons with the ambiant medium, with magnetic fields or with radiation fields. The energy of these hadrons can reach 10$^{20}$ eV. Observationally, according to Fig. \ref{Sanders}, thermal emission dominates over non-thermal emission even in radio-loud AGN (of course, except when the non-thermal emission is relativistically amplified because the jet is directed in our line of sight). 

Several mechanisms have been invoked for the jet generation and collimation. Everybody agrees that a magnetic field is indispensable to explain extended jet acceleration. One invokes centrifugally driven flows, but the presently favored mechanism is that of Blandford \& Znajek \cite{Blan}, where rotational energy is extracted from the BH via an electric field connecting the disc to the BH. The model allows  to understand  why radio-loud AGN prefer elliptical galaxies, which are thought to be formed often by the fusion of two massive spiral galaxies. During this process, one would expect the two SMBHs to merge, leading to a rapidly spinning black hole. This scenario however faces non-solved questions.  Is it sure that the two black holes will merge together, or will not one be ejected by sling-shot effect? Why do radio-quiet AGN (like MCG-6-30-15 mentioned before) have a spin? How to explain that a majority of quasars should be spinning to account for the efficiency required by the evolution of quasars at high redshifts, while radio-loud quasars are very rare precisely at high redshifts? etc...
 
\section{Weak accretors: Low Luminosity AGN, $\dot{m} \le 0.01$}

The nucleus of the giant FRI radio-galaxy M87, which displays a radio, optical, and X-ray relativistic jet at all scales from 100 $R_{\rm G}$ up to kiloparsecs, contains a 3 10$^9$ M$_{\odot}$ black hole. It has a bolometric luminosity much fainter than the Eddington luminosity, and at least two orders of magnitude smaller than the Bondi accretion rate from the hot gas surrounding the BH \footnote{It is the accretion rate of a spherical flow with a radial velocity equal to the sound velocity.} (Fabian \& Rees \cite{FabianRees}, Di Matteo et al. \cite{DiMatteo}). M87 is a LLAGN, as are all FRI galaxies, as well as radio-quiet LINERs. These objects have the same observational properties, notably the absence of a Big Blue Bump, meaning that they have no standard accretion discs. They share other properties: they generally show an intense and variable X-ray emission, their nucleus is radio-loud (this is not to be confused with the radio-loudness of the whole galaxy, see later), their spectral distribution peaks generally in the infrared, the broad FeK line is absent or weak, emission lines are generally double-peaked (Ho \cite{Ho99}, \cite{Ho02}).  

 From a theoretical point of view, computations performed in the framework of the $\alpha$-prescription show that the flow has a low density and is optically thin, owing to the small accretion rate. Like the corona discussed previously,  the disc is supported by gas
pressure and is
geometrically thick, with a large radial velocity. Therefore 
 an important fraction of the 
gravitational energy is advected towards the BH and simply swallowed. These discs have thus a low efficiency, like those accreting at a supercritical rate. 
They 
are called ``Advection 
Dominated Accretion Flows" (ADAFs, Narayan \& Yi \cite{Narayan}) \footnote{Actually the ADAF solution is valid for $\dot{m} \le 10 \alpha^2$.}. The flow is hot only close to the BH, and is probably replaced  at larger radii by a thin disc, whose UV radiation allows the hot flow to cool by Inverse Compton scattering combined with thermal bremsstrahlung. ADAFs are challenged by several other models like ``Convection dominated Accretion Flows" (CDAFs, Quataert \cite{Quataert}), ``Adiabatic Inflow-Outflow Solution" (ADIOS, Blandford \& Begelman \cite{BB99}), all being now referred to as ``Radiatively Inefficient Accretion Flows" (RIAFs).  In all models, the X-ray luminosity competes with the optical one, there is no ``Big Blue Bump", and the whole luminosity is very small compared to the luminosity that one would expect from the amount of gas available for accretion. 

However things are not that simple. First, simulations show that only a fraction of the gas supply falls on the BH, the rest being released in a bipolar outflow, owing to the high
thermal energy content of the hot gas (and probably also because the thick structure facilitates
the collimation of the jet, like at the other extreme the thick discs in strong accretors). Second, ADAF models underpredict systematically the low-frequency radio flux of LLAGN, and another component, presumably from a jet, should be added to that emitted by the disc.  This is  clear when the LLAGN is a FRI radio-galaxy like M87, as the jet is prominent. But is it also the case for radio-quiet LLAGN? 

\begin{figure}
\begin{center}
%\epsfxsize=10cm 
%\epsfbox{SagA.eps}
\includegraphics[width=8.0cm]{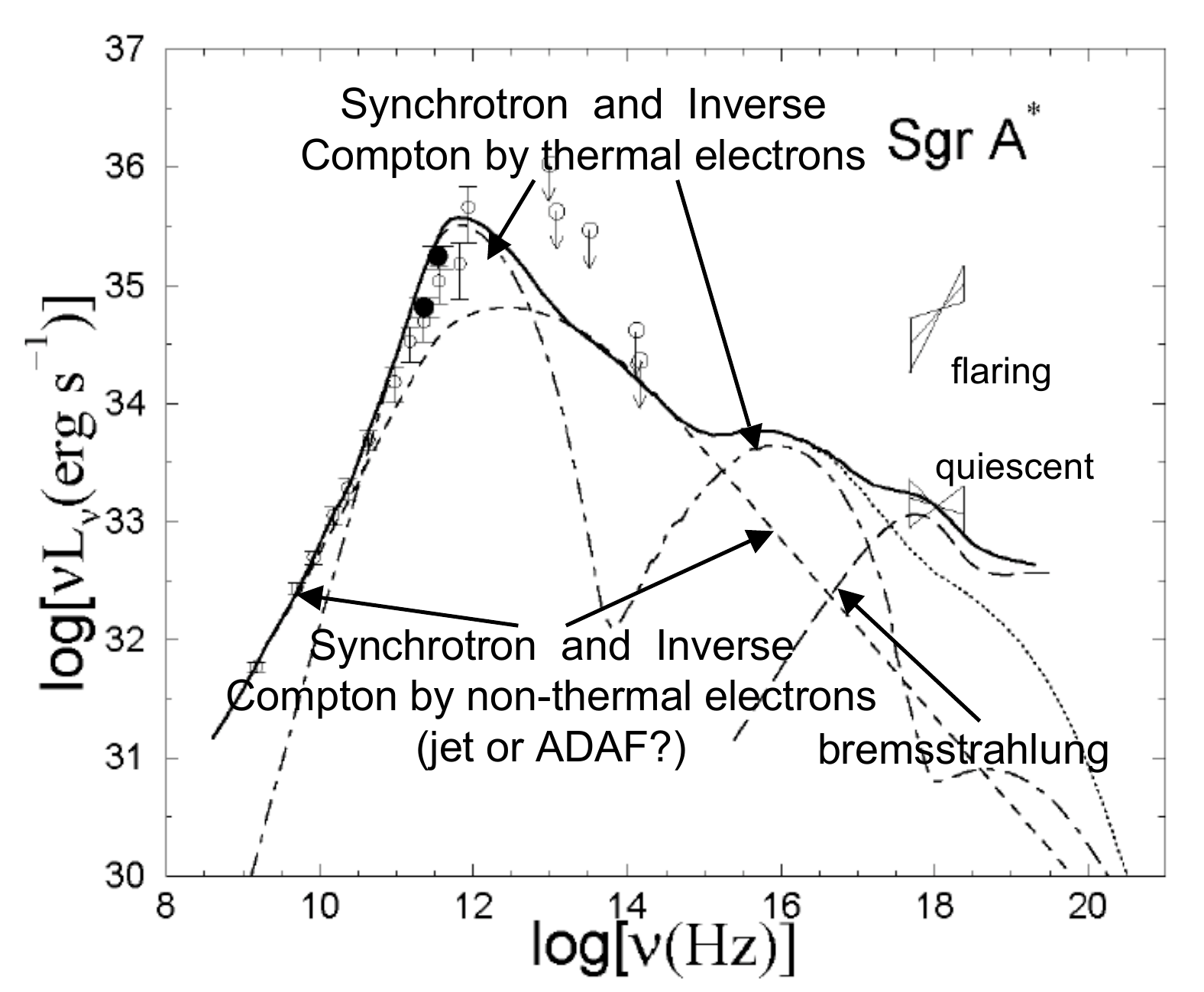}
\caption{Spectral energy distribution of Sgr A* fitted by an ADAF plus a synchrotron and an Inverse Compton emission, from Yuan et al. 2006.}
\label{SagA}
\end{center}
\end{figure}

Several LLAGN were modeled successfully with an ADAF plus a jet, in particular NGC 4258, a weak Seyfert for which we have precise data, thanks to observations of water masers - mass, distance, broad band spectrum. In this object, Yuan \cite{Yuan02} find that {\it the emission is entirely dominated by the jet}.  
The best known object in this context is Sagittarius A* (Sgr A*), the radio-source associated with the black hole at the center of our Galaxy. The BH mass is 3.5 10$^6$ M$_{\odot}$, and the accretion rate (due to outflows of hot stars and to molecular clouds) would correspond to a luminosity of the order 10$^{41}$ ergs s$^{-1}$, while it is observed to be only $\sim 10^{36}$ ergs s$^{-1}$, implying an efficiency $\epsilon\sim 10^{-6}$. Fig. \ref{SagA} gives a fit of the spectrum of Sgr A* in a quiescent phase by an ADAF (Yuan et al. \cite{Yuan06}). The authors were obliged to add  the contribution of synchrotron radiation in the radio range, and correlatively Inverse Compton in the gamma-ray range.
It is also worth mentioning that an alternative model is presently developed:  a geometrically thin disc coupled with a magnetized  jet. It gives also a good fit to the data of M87 and of Sgr A* and seems physically justified (Jolley \& Kuncic \cite{Jo07}, \cite{Jo08}). In both objects, the jet would extract a large fraction of the mass and the accretion power, and the real accretion rate onto the BH would be strongly reduced (consequently the efficiency would be almost ``normal").

All this raises the question of whether {\it all LLAGN power jets and produce non-thermal radiation, and at which level}.

\begin{figure}
\begin{center}
\includegraphics[width=8.0cm]{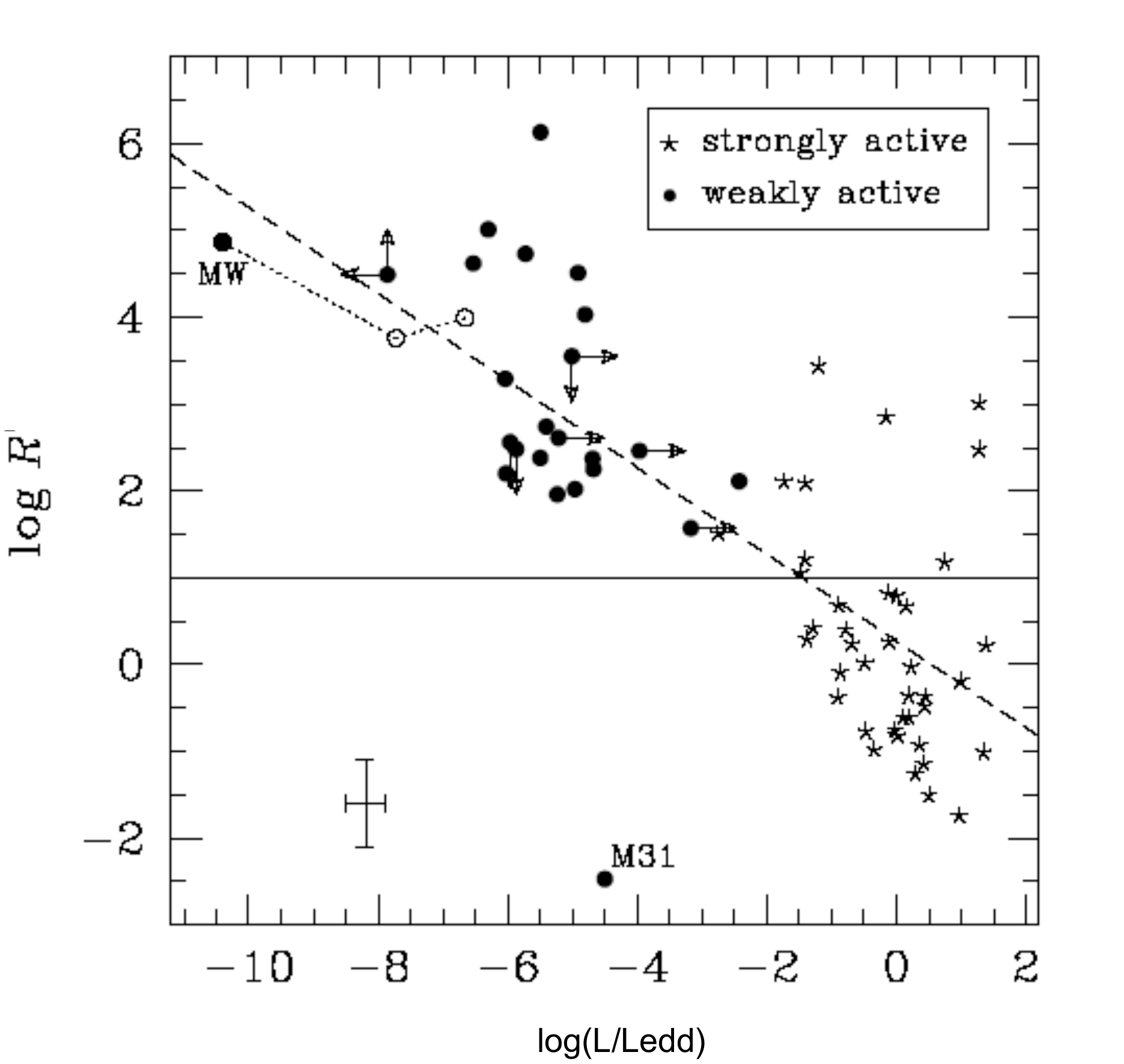}
%\epsfxsize=15cm 
%\epsfbox{Ho0208.eps}
\caption{Core radio luminosity at 6cm to optical  luminosity ratio, $R$,  versus the Eddington ratio, from Ho 2002. The three points connected by a dotted line are at the center of the Milky Way (MW). The point with the larger value of $R$ corresponds to Sgr A*, i.e. to the supermassive BH.  }
\label{Ho02}
\end{center}
\end{figure}

When distinguishing previously between radio-loud and radio-quiet AGN (Fig. \ref{sikora}), we included the emission of radio-sources at large distance from the nucleus, while the optical luminosity was restricted to the nucleus. It was reasonable since the nucleus is at the origin of the big radio-sources. But if we want to know whether there are jets and relativistic electrons {\it inside} the nucleus, we have to consider only the core radio luminosity. Fig. \ref{Ho02} shows the ratio of the {\it radio core} to optical luminosity as a function of the Eddington ratio, for a sample of nearby galactic nuclei, including LLAGN, Seyfert and quasars. We see that the two parallel sequences have disappeared for LLAGN, meaning that {\it the core radio luminosities of radio-loud and radio-quiet objects are similar}. Moreover, nuclear radio loudness is a strongly decreasing function of the Eddington ratio. Note that LLAGN are often nuclei of early type galaxies with large BH masses, on the order 10$^{9}$ M$_{\odot}$. 

The empirical connection between LLAGNs and jets is thus
established from radio observations. Not only the radio emission is measured, but in many
cases, the jets themselves can be seen directly from VLBI-scale radio images.  One should determine the amount of energy transported by these jets. Relying on models and on analogies with stellar black holes in the low state and with radio-loud AGN, one can determine the kinetic power injected in the jets: it is found to be at least of the same order as the bolometric luminosity. Note that, if the efficiency is not small as proposed by Jolley and Kuncic, the outflowing rate could be much larger than the accretion rate on the BH, and close to the Bondi rate. 

If jets are produced by spinning BHs according to the Blandford-Znajek mechanism, the acceleration mechanism for UHECRs is likely to take place in the generated external current, and the $emf$  is thus equal to 4 10$^{20} B_4M_9$V at the gravitational radius, where $B_4=B/10^4{\rm G}$ and  $M_9=M/10^9$M$_\odot$. Considering the BH masses of nearby galaxies, Boldt \& Ghosh \cite{Boldt} determined the maximum value of $B$, assuming equilibrium between the gas pressure for the maximum accretion rate and the magnetic pressure, and they obtained a typical maximum $emf$ of a few 10$^{20}$ V.
A condition is that the
radiation background be extremely
low in order to avoid losses due to pion and pair production. This is exactly what happens with RIAFs. 
Detailed computations for this kind of models should obviously be more elaborated, but it is at least not completely utopian. 

\begin{figure}
\begin{center}
\includegraphics[width=8.0cm]{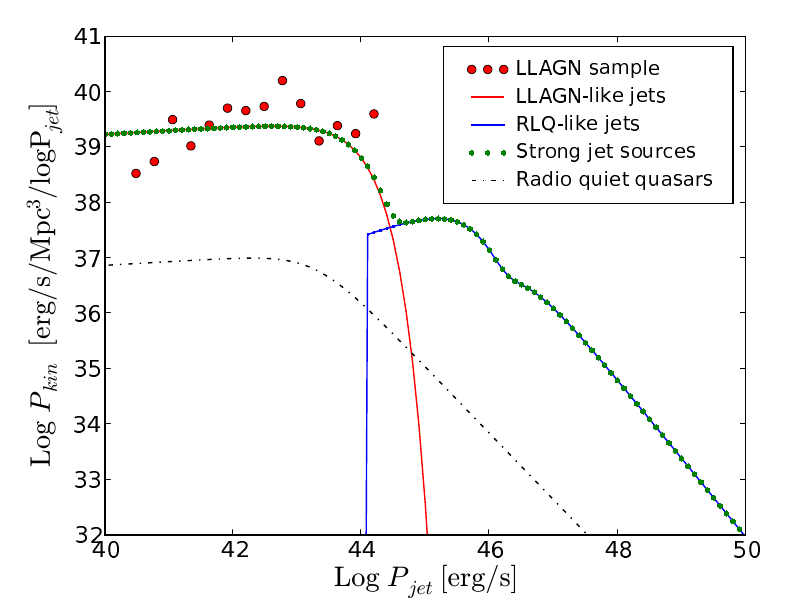}
%\epsfxsize=10cm 
%\epsfbox{Kording08.eps}
\caption{The kinetic luminosity function of jets as a function of jet's power at redshift zero, after K\"ording et al 2008. It is dominated by LLAGN. }
\label{Kording08}
\end{center}
\end{figure}

Another proof of the strong influence of compact jets of LLAGN is given by the comparison of the total power generated by these jets in the intergalactic medium, compared to that of large scale jets of FRI and FRII radio-galaxies or radio-loud quasars. 
We have seen (Fig. \ref{Sanders}) that among luminous AGN, {\it only radio-loud AGN produce non-thermal photons, and these photons do not dominate the radiative power}. The sample of Fig. \ref{sikora} being not statistically complete, one does not realize that 
only a small fraction of AGN are radio-loud: 10$\%$ in the local Universe, and less at high redshift (McLure \& Jarvis \cite{McLure}). Thus {\it the proportion of radio-loud AGN with respect to all LLAGN is of the order of 0.2$\%$}. 
We are led to  conclude that if only radio-loud AGN emit non-thermal photons, this would be a problem because they are very rare objects \footnote{But it is not the case when trying to find which objects give rise to the bulk of non-thermal emission and relativistic particles in the Universe. In particular, those directed towards us (the blazars) remain the best {\it individual} locations to look for high energy particles and  photons.}. And indeed K\"ording et al. \cite{Kording} showed that the total kinetic energy injected in the intergalactic medium by jets of LLAGN dominates by at least one order of magnitude on that of radio-loud AGN (Fig. \ref{Kording08}). One does not know whether this result can be applied to the generation of high energy particles,  but it is most likely. 

\section{Conclusion}

We have shown that luminous radio-quiet AGN (Seyfert nuclei and quasars) do not emit non-thermal radiation, and that they produce winds and not relativistic jets. On the contrary, jets are produced systematically in low luminosity AGN (LLAGN), whose bolometric luminosity is dominated by non-thermal radiation. The jets in radio-loud quasars and powerful (FRII) radio-galaxies are individually radiating more non-thermal radiation than those in LLAGN, but the conditions for confining and accelerating UHECRs can probably be met as easily in compact jets working with the Blandford-Znajek dynamo. Moreover, the latter are almost three orders of magnitude more numerous, as they constitute about 40$\%$ of ordinary galaxies: in all, they provide at least one order of magnitude more energy than jets of radio-loud AGN, and probably as much UHECRs.

\end{document}